\documentclass[fleqn,10pt]{wlscirep}
\usepackage{color,graphicx}

\title{Energy correlations of photon pairs generated by a silicon microring resonator probed by Stimulated Four Wave Mixing.}

\author[1]{Davide Grassani}
\author[1]{Angelica Simbula}
\author[1]{Stefano Pirotta}
\author[1]{Matteo Galli}
\author[1]{Matteo Menotti}
\author[2]{Nicholas C. Harris}
\author[3]{Tom Baehr-Jones}
\author[3]{Michael Hochberg}
\author[4]{Christophe Galland}
\author[1]{Marco Liscidini}
\author[5,*]{Daniele Bajoni}

\affil[1]{Dipartimento di Fisica, Universit\`{a} di Pavia, via Bassi 6, 27100 Pavia, Italy}
\affil[2]{Department of Electrical Engineering and Computer Science, Massachusetts Institute of Technology, 77 Massachusetts Avenue, Cambridge, MA 02139, USA}
\affil[3]{Coriant Advanced Technology Group, 1415 West Diehl Road, Naperville, IL 60563, United States}
\affil[4]{Ecole Polytechnique Fédérale de Lausanne, SB-LPQM, CH-1015 Lausanne, Switzerland}
\affil[5]{Dipartimento di Ingegneria Industriale e dell'Informazione, Universit\`{a} di Pavia, via Ferrata 1, 27100 Pavia, Italy}

\affil[*]{daniele.bajoni@unipv.it}

\begin{abstract}
Compact silicon integrated devices, such as micro-ring resonators, have recently been demonstrated as efficient sources of quantum correlated photon pairs. The mass production of integrated devices demands the implementation of fast and reliable techniques to monitor the device performances. In the case of time-energy correlations, this is particularly challenging, as it requires high spectral resolution that is not currently achievable in coincidence measurements. Here we reconstruct the joint spectral density of photons pairs generated by spontaneous four-wave mixing in a silicon ring resonator by studying the corresponding stimulated process, namely stimulated four wave mixing. We show that this approach, featuring high spectral resolution and short measurement times, allows one to discriminate between nearly-uncorrelated and highly-correlated photon pairs.
 
\end{abstract}
\begin{document}

\flushbottom
\maketitle
%
%
\thispagestyle{empty}


\section*{Introduction}

Since the seminal work of Politi et al. \cite{Politi2008}, there has been tremendous progress in the field of integrated quantum optics. Integrated photonic structures fabricated via femto-second laser writing in glass, electron-beam lithography, and photolithography in semiconductor materials are arguably the most promising route towards the successful implementation of quantum information and communication protocols \cite{Peruzzo2010,Meany:OpEx:12,Crespi2013,Silverstone2014}. Photonic integrated circuits  \cite{Streshinsky:13} may indeed overcome the practical limitations of bulk optics in terms of system scalability and flexibility. 

One of the main issues that still hinders the development of quantum optics is the efficient and on-demand generation of non-classical states of light. In most of the above-mentioned experiments the generation of light is achieved by means of bulky nonlinear crystals without taking any advantage of integrated optical elements. 
Still, there have been also several proposals for the generation of photon-pairs in integrated devices based on either spontaneous parametric down conversion (SPDC) or spontaneous four-wave mixing (SFWM) in $\chi^2$ and  $\chi^3$ materials, respectively \cite{Azzini2012OE,Orieux2013,Silverstone2014}. 
As for passive optical components, it is clear that the use of integrated devices can be of great help to enhance the generation efficiency and tailor the properties of the generated light. The goal is the integration of hundreds of these sources on a single chip, for example in the view of multiplexing them to achieve on-demand sources of non-classical light \cite{Collins:2013eu,Harris:PRA:2015}

Among the different strategies to generate photon pairs in integrated optical components, silicon and silicon compatible ring resonators \cite{Azzini2012OL,Reimer2014,Ramelow2015} are particularly appealing: they are CMOS compatible, have minimal footprint, and feature a very high spectral brightness. It has been shown that the nonlinear response of the bulk materials can be enhanced by more than 7 orders of magnitude by light confinement in silicon rings having a radius of a few tens of micrometers and quality factors of about $10^4-10^5$ \cite{Azzini2012OL}. It has been shown that for a pump power of about $1$ mW, a silicon microring can generate up to $10^7$ photon pairs/s within a typical spectral bandwidth of hundreds of pm.  This is particularly suitable for quantum communications over the telecom infrastructure, in which low pump power is desirable and spectral bandwidth is limited. It has also been shown that the generated pairs are time-energy entangled \cite{Grassani:Optica:2015}.  Finally, it has been suggested that by only changing the coherence properties of the pump excitation, one can in principle control the spectral correlations of the generated pairs, ranging from nearly uncorrelated to highly correlated photons \cite{Helt2010}. In the case of time-energy correlations, highly spectrally correlated photons result to be in an entangled state, while indistinguishable heralded single photons can be obtained from spectrally uncorrelated pairs, without the need for additional filtering stages \cite{Spring:13}. 

This capability to produce non-classical states of light for different applications, along with the possibility of mass production of integrated optical circuits, suggests the development of fast and reliable techniques to quickly characterise the quantum properties of these devices. This is a challenging task, as it requires the reconstruction of the biphoton wavefunction, which describes all the properties of the generated pairs, in the energy Hilbert space. This is particularly demanding for ring resonators, given their narrow generation bandwidth. State-of-the-art techniques based on coincidence measurements do not have sufficient resolution to investigate photon spectral correlations of pairs generated in these devices \cite{Avenhaus:OpLett:09}. 

Another possible approach \cite{Liscidini:PRL:2013} to study the generation of photon pairs by parametric fluorescence (either SPDC or SFWM) is exploiting the corresponding stimulated nonlinear process, difference frequency generation (DFG) or four-wave mixing (FWM), respectively \cite{Azzini2012OL}. The stimulated process is intrinsically much stronger and yields very high signal-to-noise ratios, leading to a fast and highly resolved two-photon state characterisation \cite{Eckstein:LasPhotRev:2014,Fang:Optica:2014,Rozema:Optica:2015,Kumar2014,Silverstone2015}.

\section*{Results}

\subsection*{Sample and Characterization}

\begin{figure}[ht]
\centering
\includegraphics[width=\linewidth]{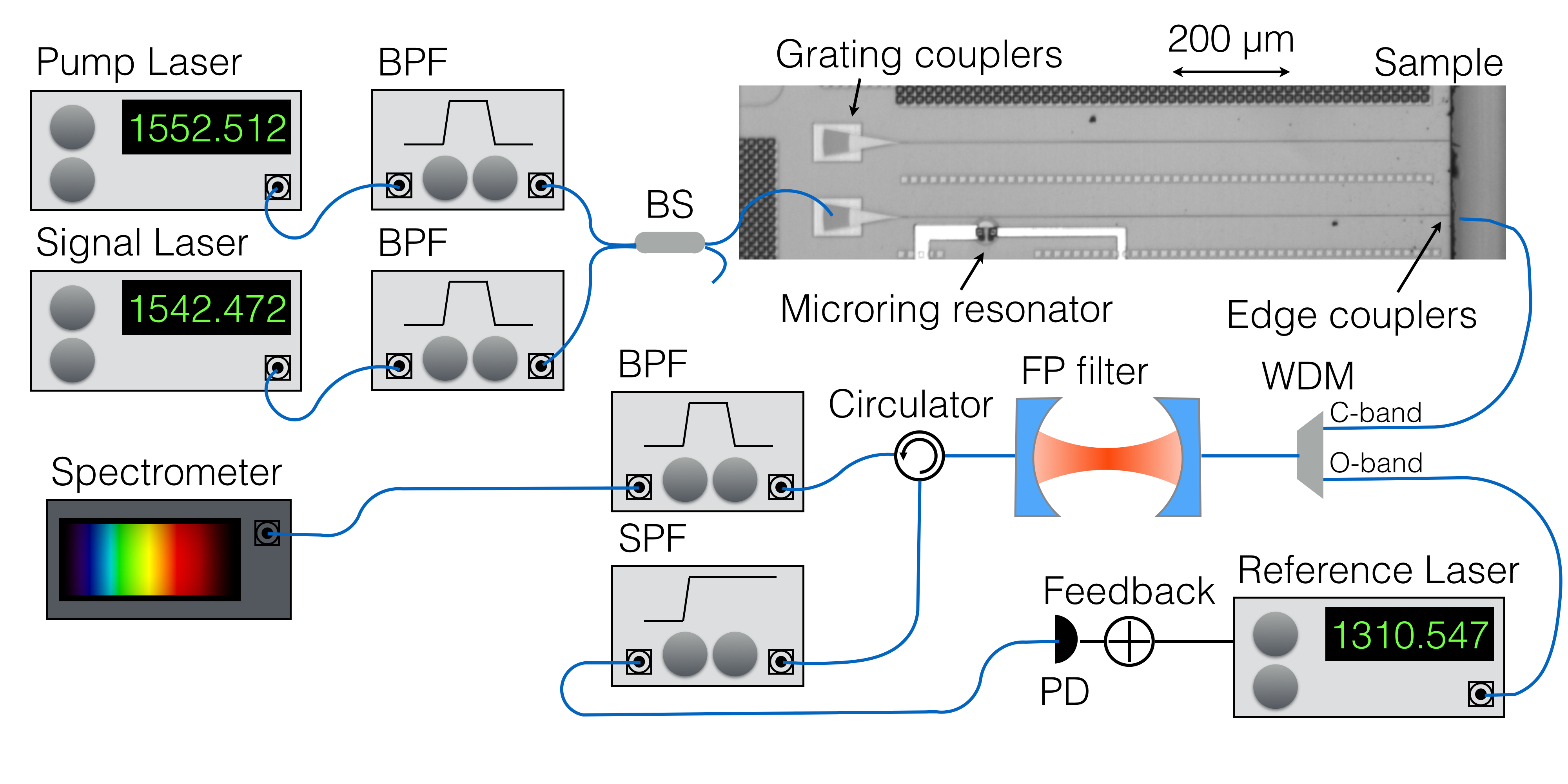}
\caption{Schematic representation of the experimental set-up. Two lasers, spectrally cleaned by Band Pass Filters (BPF) are combined in a beamsplitter (BS) and coupled to the sample. An optical micrograph of the sample is shown in the picture. The scale-bar refers to the sample only, the rest of the schematics is not in scale. A custom built Fabry Perot interferometer (FP filter) is used to achieve spectral resolution on the generated idler resonance. The FP filter position is controlled and stabilized using a reference laser in the telecom O-band and an active feedback loop. The output from the FP filter is fed to a spectrophotometer coupled to a CCD camera. PD stands for photodiode and SPF stands for Short Pass Filter.}
\label{Figsetup}
\end{figure}

Our sample is fabricated using a standard CMOS silicon photonics process featuring high yield and reproducibility (See Supplementary information for details). It consists of a silicon ring resonator with a radius $R=$15 $\mu$m, side coupled to a 500 nm wide $\times$ 220 nm high silicon bus waveguide; the entire structure is embedded in silica. A photograph of the sample is shown in Fig. \ref{Figsetup}. Light is coupled into the sample via a grating coupler \cite{He:2013hx} and out via an edge coupler. 

\begin{figure}
\centering
\includegraphics[width= 0.5\linewidth]{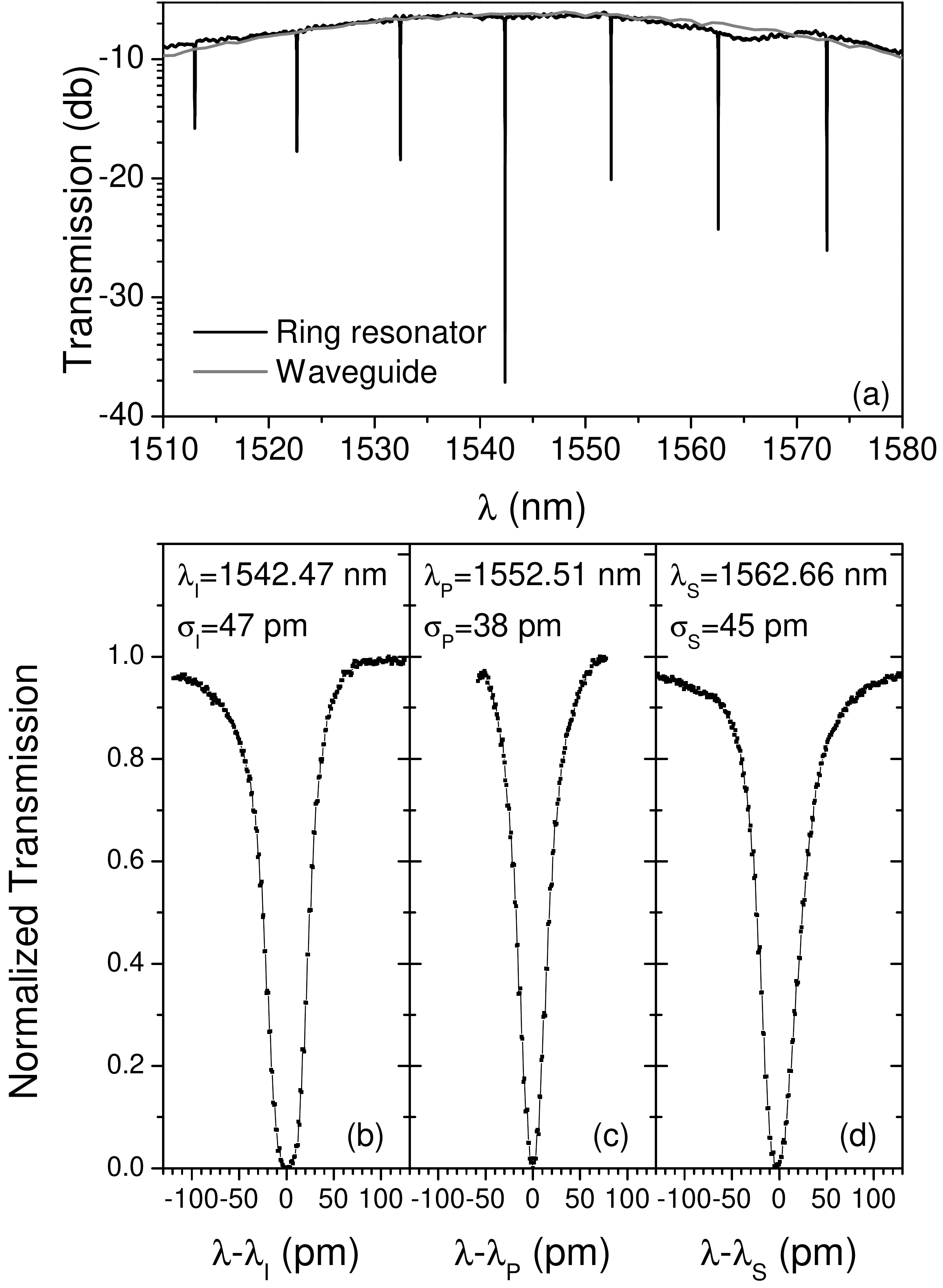}
\caption{Transmission spectrum from the sample.(a) Broad, low resolution (5 pm) spectrum of the ring resonator and the grating coupler. (b) (c) and (d) Full resolution transmission spectra of the idler, pump and signal resonances respectively.}
\label{FigT}
\end{figure}

In Fig. \ref{FigT} (a) we show the transmission spectrum of the structure under investigation along with that of a reference waveguide (i.e. without the ring resonator). In both cases the bell shape of the transmission is due to the spectral response of the grating coupler \cite{He:2013hx}. The spectral resolution is about $5$ pm thus, although several resonances are clearly visible, these are not completely resolved.  In Fig. \ref{FigT} (b) (c) and (d) we plot the detailed transmission spectra of three adjacent resonances, which are label signal, pump and idler, for they are used in our SFWM and FWM experiments. These spectra have resolutions of $2$ pm and show that the ring resonator is essentially in the critical coupling condition, in which the pump transmission ideally drops to zero. The three Lorentzian resonances have similar full width at half maxima ($\delta\lambda_i$) corresponding to a Q factor of about 40000. In particular, for the pump resonance, we have $Q_P=40800\pm2000$, equivalent to a coherence time (corresponding to the dwelling time inside the resonator) $\tau_p=1/\Delta\omega\sim33\pm2$ ps.

\subsection*{Spontaneous Four Wave Mixing}

\begin{figure}
\centering
\includegraphics[width= 0.7\linewidth]{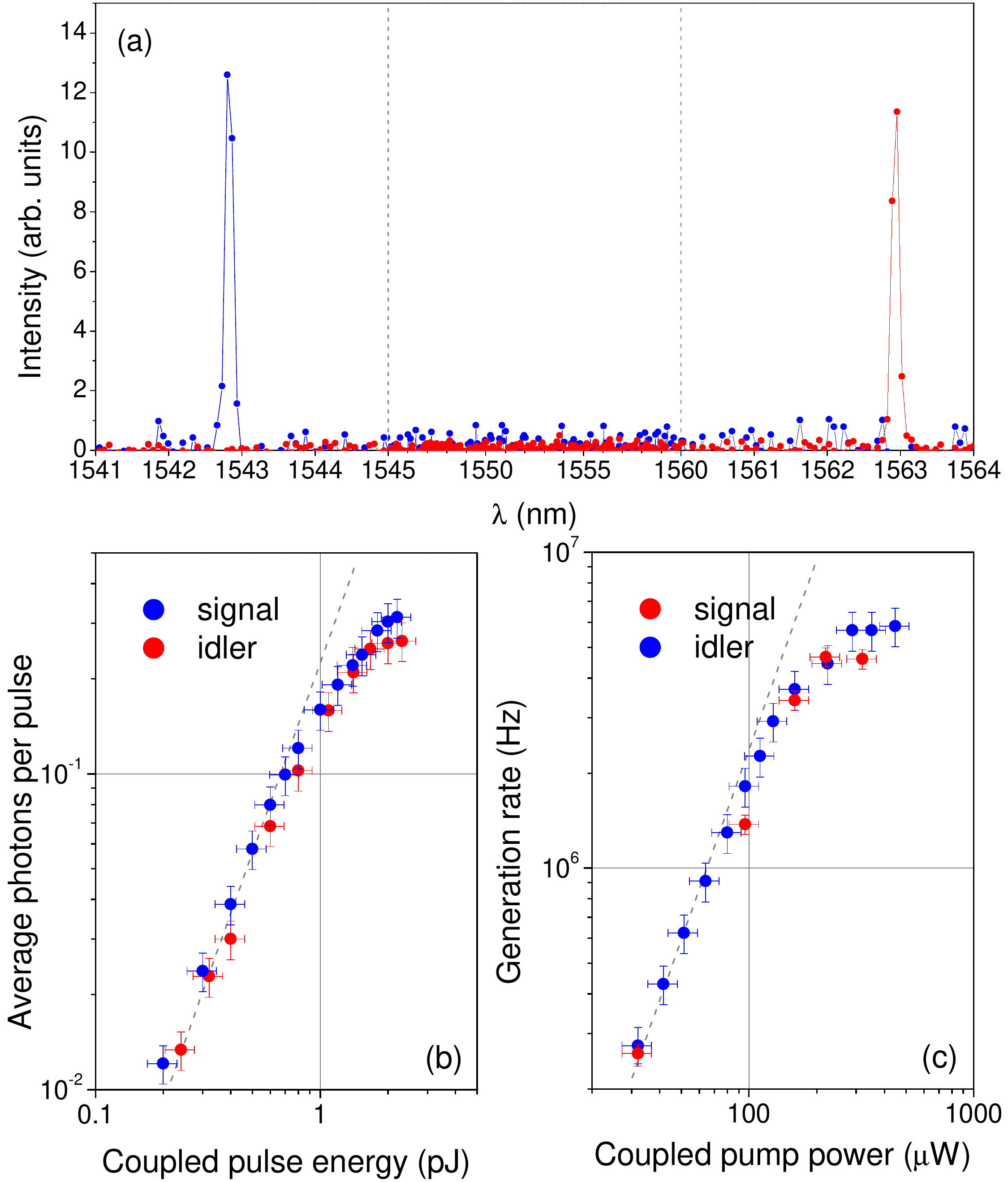}
\caption{(a) An example of spontaneous four wave mixing spectrum obtained with pulsed pumping (0.8 pJ per pulse), showing the generated signal and idler peaks and the full rejection of the pump. (b) Dependence of the number of generated photons per pulse on the pump's energy per pulse. (c) Dependence of the generation rates on the pump power for the cw case. Dashed lines are guides to the eye proportional to the square of the pump energy per pulse (in panel (b)) and pump power (in panel (c)).}
\label{FigFWM}
\end{figure}

When the sample is pumped using a pulsed (Pritel FFL) laser tuned at $\lambda_P$, having an energy of 0.8 pJ per pulse and a repetition rate of 10 MHz, the spectrum taken with a  liquid nitrogen  (LN2) cooled CCD camera shows emission of photons at the signal and idler resonances due to SFWM (see  Fig. \ref{FigFWM} (a)).  Here, external filters are used to remove the intense residue of pump light and separate signal from idler photons leading to a negligible background (see Fig. \ref{Figsetup}). It should be noticed that on-chip filtering and routing has also been demonstrated on the same sample \cite{Harris:PRX:2014}. 

In Fig. \ref{FigFWM} (b) we show the average number of generated photons versus the pulse energy. The same experiment has been repeated by pumping the sample with a continuous wave (cw) (Santec TSL 510) laser, and photon generation rates are shown in Fig. \ref{FigFWM} (c) as a function of the pump power. For both the pulsed and cw experiments the quadratic scaling of the generation rate with the pump power confirms that in this case linear parasitic processes, like spontaneous Raman scattering and photoluminescence, are negligible. The curve saturation at high pump powers is caused by the onset of the ring bistability due to two photon absorption \cite{Azzini2012OL}. Pump powers are estimated by measuring the power at the sample input and correcting for the coupling losses from the grating coupler (5 dB \cite{Harris:PRX:2014}). The total length of the waveguide is less than a mm, thus propagation losses (of about 3 dB/cm) are negligible. We estimate the internal generation rate by measuring the signal and idler intensity at the chip output and taking into account for the coupling losses \cite{Grassani:Optica:2015}. The generation rate inside the ring is comparable to what was reported in similar structures \cite{Grassani:Optica:2015,Harris:PRX:2014}.

\subsection*{Joint Spectral Density Measurements}

To characterise the spectral correlations of the generated photons, we reconstruct the joint spectral density (JSD), which is the square modulus of the biphoton wavefunction. In practice, the JSD gives the wavelength of the signal (idler), as a function of the wavelength of the idler (signal). In order to reconstruct the JSD in a coincidence measurement on spontaneously generated photons, for each signal wavelength inside the signal resonance all the idler wavelengths inside the idler resonance have to be recorded. However, the JSD can be obtained also by exploiting the far more efficient stimulated process (FWM) \cite{Fang:Optica:2014}. In this case, the spectral resolution on the signal resonance is given by that of the cw seed, while idler photons have to be filtered with a tunable band pass filter (BPF) narrow enough to resolve the idler resonance, as in the set-up schematized in Fig. \ref{Figsetup}.

Here, the sample is pumped either with a pulsed or cw laser  as in SFWM but, in addition, a second tunable cw laser (Santec TSL 510), whose wavelength can be controlled with an accuracy of about 2 pm, is employed to stimulate the generation of pairs. Band Pass Filters (BPFs) are used after each laser to reject the broadband background arising from amplified spontaneous emission. The signal and the pump lasers are combined using a polarization maintaining beam splitter, with one of the beam splitter's outputs being coupled to the sample. Light is collected at the sample output through a lensed fiber, and a home-made Fabry-P\'erot filter is used to analyze the generated idler beam. This filter is actively stabilized and has a spectral resolution of about 5 pm (see Supplementary information). The output of the Fabry-P\'erot is then detected via a spectrometer coupled to a liquid nitrogen cooled CCD camera.

\begin{figure*}
\begin{center}
\includegraphics[width= \textwidth]{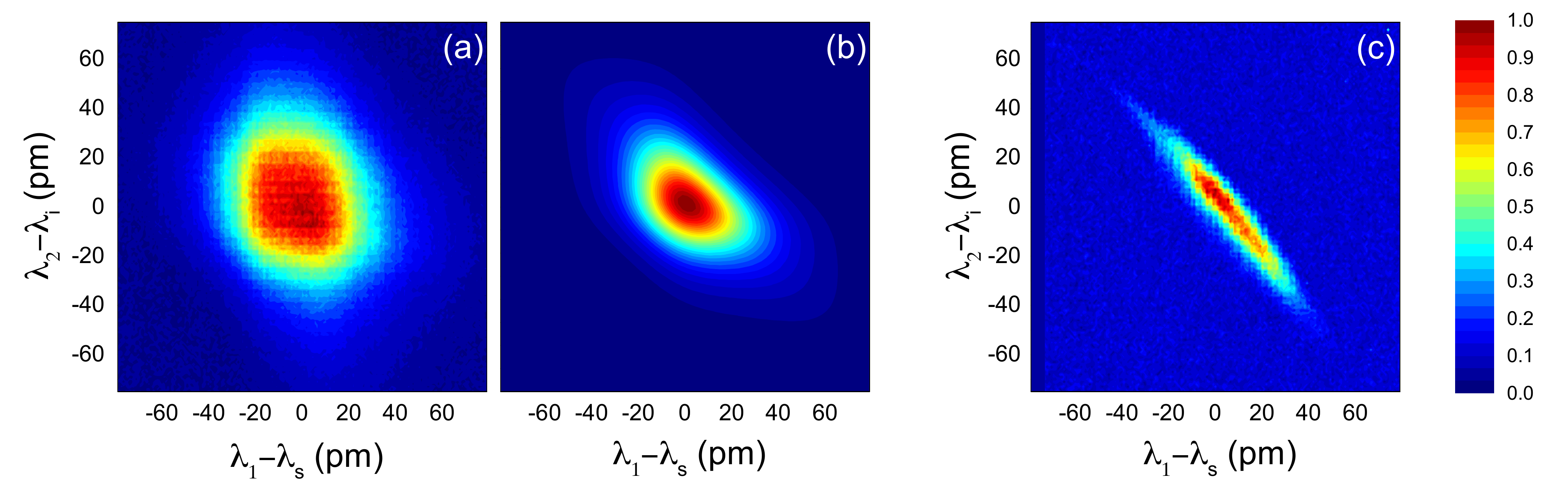}
\caption{(a) Joint spectral density measured  under pulsed pumping, to be compared with the calculated Joint Spectral Density in (b). (c) Joint spectral density measured  under cw pumping.}
\label{FigJSD}
\end{center}
\end{figure*}

First we consider the pulsed laser case, in which the pump BPF sets the pump line width to $90$ pm, corresponding to a pulse duration of 14 ps while the pulse energy is 0.8 pJ. The pulse linewidth is larger than the linewidth of the ring resonator mode around 1552 nm, and thus it is filtered by the resonance (as accounted for in the theory). In Fig. \ref{FigJSD} (a)  and (b) we show the measured JSD along with the theoretical result obtained by considering in our theory \cite{Helt2010} the experimental sample parameters  (ring radius R=$15$ $\mu$m, n$_\mathrm{eff}$=$2.54$, v$_\mathrm{g}$=$116$ $\mu$m/ps, group velocity dispersion GVD=$1.84$ $\mu$m$^2$/ps). The theoretical simulation does not include a slight ($\sim$ 10\%) broadening of the resonances due to two photon absorption under pulsed pumping.

Assuming a pure state, the determination of the full biphoton wavefunction allows one to calculate the Schmidt number $K$, where a separable state corresponds to K=1, while an entangled state is characterised by $K>1$.  From our theoretical model we found $K$=1.09,  which indicates nearly uncorrelated photons, as it is expected in this system when the pump pulse duration is equal or shorter than the dwelling time of the photons in the ring  \cite{Helt2010}. The weak energy correlation is also visible from the JSD, which allows one to determine a lower bound $K_{bound}$ for the Schmidt number. From the data reported in Fig. \ref{FigJSD} (a) we derive $K_{bound}$=1.03 with and uncertainty of $0.1$, which well compares with the expected theoretical value. The measured value is lower than the calculated Schmidt number (although within the uncertainty, see Supplementary Information) due to the finite spectral resolution in the experiment.

In a second experiment, we consider a cw pump for which the generation of entangled photon pairs is expected and has been recently demonstrated \cite{Grassani:Optica:2015}. The pump power inside the ring is 80 $\mu$W. We show the measured JSD in Fig. \ref{FigJSD} (c). The elongated shape of the JSD indicates spectral correlations, and the lower bound of the Schmidt number calculated from the data is indeed $K_{bound}$=3.93, a value limited by our experimental resolution (and in particular of the Fabry Perot filter, see Supplemental Information). The coherence time of the pump is about 1 $\mu$s, which in this ring corresponds to a theoretical Schmidt number $K$=37038. It should be noticed that, despite our lower bound being much smaller than the expected Schmidt number, we are still able to discriminate between the generation of correlated and nearly uncorrelated photons. In the present case this would not be possible with state-of-the-art techniques based on coincidence measurements, which have a resolution of hundreds of pm, (i.e. the typical pixel is larger than the whole Fig. \ref{FigJSD} (a) or (c)). Finally, we stress that the data acquisition for Fig. \ref{FigJSD} (a) and (c) took about one hour of total integration time.

\section*{Discussion}

We have experimentally demonstrated that a microring resonator integrated on a silicon chip can be driven to emit nearly uncorrelated or time-energy entangled photon pairs depending on the pump 
. This has been done by exploiting FWM to directly and rapidly reconstruct the JSD of the generated photon pairs by the spontaneous process.

Thanks to the high signal-to-noise ratio achievable in the stimulated process, we could spectrally filter the generated photons with a band pass filter of only few picometers width and avoid the use of single photon detectors.
The resolution we achieved is less than 10 pm$^2$, more than two orders of magnitude better than state-of-the art measurements based on single photon detection. This resolution could be further improved by the use of state of the art Optical Spectrum Analyzers with MHz resolution, instead of the Fabry Perot filter. 

The experimental times are also greatly reduced, and the acquisition of the data in Fig. \ref{FigJSD} (a) or (c) took about one hour. In a typical coincidence experiment several minutes of integration would be needed for each point in the JSD diagrams.  We believe this results are particularly interesting for the development of fast and reliable approaches to characterise on-chip sources of non-classical states of light. 

\section*{Methods}

The fabrication of the sample was carried out by the OpSIS foundry service in a CMOS-compatible process at the Institute of Microelectronics (IME) in Singapore. The OpSIS  project was based at the University of Delaware. The starting material is a Silicon-on-Insulator(SOI) wafer from Soitec with a 220nm-thick top crystalline silicon layer and a 2 mm buried oxide layer (BOX).
The photonic circuits are defined by means of litography and reactive ion etching. The photoresist material is deposited on top of the sample, and 248 nm UV-photolitography is used in order to transfer the circuit design from the mask to the photoresist. The optical devices are then realized by means of anisotropic dry etching with different depths: grating couplers are obtained from a shallow etch of 60nm, while 130 nm and 220 nm depths are used respectively for rib and ridge waveguides.

From the study of transmission properties of the sample, the measured losses were 2.7$\pm$0.06 dB/cm for the ridge waveguide and 1.5$\pm$0.6 dB/cm for the rib waveguide. \cite{Galland:SPIE:2013}

More specifically, the device used in this experiment is composed by a grating coupler, a 950$\mu$m long waveguide, and a microring with radius of 15$\mu$m, critically coupled to the waveguide. 
On one side of the waveguide there is the grating coupler, used as input, in which light is injected by means of an almost vertical fiber array: its angle with respect to the normal is set in order to optimize transmission around 1540 nm. The other extremity of the waveguide ends in the edge of the sample, where the light coming out of the sample is collected by a tapered fibre. 


The Fabry-Pérot filter used to scan the idler resonance is made by two 90/10 fused silica beam splitters mounted on translating stages and controlled by piezo-electric actuators (see Supplementary material for further details). The output beam from the sample is sent to and collected from the Fabry-Pérot by two identical fiber-to-free-space collimators, with numerical aperure NA = 0.24, focal length = 37 mm and beam diameter about 7 mm.



\section*{Acknowledgements}

We acknowledge support from MIUR through the FIRB ``Futuro in Ricerca'' project RBFR08XMVY, from the foundation Alma Mater Ticinensis and by Fondazione Cariplo through project 2010-0523  Nanophotonics for thin-film photovoltaics.

\section*{Author contributions statement}


NCH TBJ MH and CG were involved in the design and fabrication of the sample, DG AS SP MG and DB were involved in the experimental measurements and ML and MM were involved in the theoretical part of the work. ML and DB wrote the main manuscript text, DG AS ML and MM the supplementary information. All authors reviewed the manuscript.

\section*{Additional information}
The authors declare no competing financial interests.


The corresponding author is responsible for submitting a \href{http://www.nature.com/srep/policies/index.html#competing}{competing financial interests statement} on behalf of all authors of the paper. This statement must be included in the submitted article file.




\end{document}